\documentstyle[12pt]{article}
\newcommand{\ber}{\begin{eqnarray}}
\newcommand{\eer}{\end{eqnarray}}
\newcommand{\bea}{\begin{equation}}
\newcommand{\eea}{\end{equation}}
\begin{document}
\title{SCALING FUNCTION FOR THE CRITICAL SPECIFIC HEAT IN A CONFINED GEOMETRY
: SPHERICAL LIMIT}
\author{\bf {Saugata Bhattacharyya and J. K. Bhattacharjee} \\
Department of Theoretical Physics \\
Indian Association for the Cultivation of Science \\
Jadavpur, Calcutta 700 032, India}
\date{}
\maketitle
\begin{abstract}
The scaling function for the critical specific heat is obtained
exactly for temperatures above the bulk transition temperature by
working in the spherical limit. Generalization of the function to
arbitrary $\alpha$ (the specific heat exponent), gives an
excellent account of the experimental data of Mehta and Gasparini
near the superfluid transition.\\\\\\
PACS number(s): 05.70 Jk ; 64.60.-i ; 64.60.Fr ;
\end{abstract}
\newpage
One of the primary theoretical tools for a strong coupling problem is the
spherical limit and the $\frac{1}{n}$ expansion [1-5]. In critical phenomena,
it has often provided useful insight in the past and is still one of the first 
techniques to be tried when a new kind of critical phenomenon surfaces [6-11].
A problem that has been receiving a fair amount of experimental and theoretical 
attention is that of the specific heat of liquid helium near the lambda point
in a confined geometry. This is a very effective way of studying the finite 
size effects (FSE) since by going close enough to the lambda point 
(particularly easy in a space-shuttle experiment), the correlation length 
can be made larger than the confining length if the liquid is enclosed in a
thin wafer geometry. The bulk specific heat $C_p$ diverges near the lambda 
point as $\xi^{\alpha/\nu}$ , where $\xi$ is the correlation length 
proportional to $\mid T-T_c\mid^{-\nu}$. For the finite size system, the 
specific heat remains finite at $T=T_c$ , with the value proportional to
$L^{\alpha/\nu}$ where L is the confining length. For a given L, the specific
heat changes from a $\xi^{\alpha/\nu}$ to a $L^{\alpha/\nu}$ as one approaches 
the bulk lambda point. The change from one regime to the other is described in
terms of a scaling function $f(x)$ in terms of which we can write the specific 
heat as 
\bea
C_p(\xi,L) = C_0 \xi^{\alpha/\nu} f(\xi/L) + Constant
\eea
where the function $f(x)$ has the property that $f(0)=1$ and $f(x)\sim 
x^{-\alpha/\nu}$ for $x \gg 1$ . It is the function $f(x)$ that has been the
object of several experimental and numerical investigations. In this letter,
we determine $f(x)$ in the spherical limit. We also show how the result can
be put to practical use.
\par
We consider a n-component Ginzburg-Landau model with the free energy 
functional
\bea
F = \int d^{D}r\:\: [ \frac{m^2}{2}\sum_{i=1}^{n} \phi_i^2 + \frac{1}{2}
\:\sum_{i=1}^{n}\:(\nabla\phi_i)^2 + \lambda\:(\sum_{i=1}^{n} \phi_i^2)^2\:]
\eea
where $m^2 \sim (T-T_0)$ , $T_0$ being a transition temperature and the    
coupling constant $\lambda$ is $O(n^{-1})$ so that the quartic term remains 
the same order as the quadratic term when $n \rightarrow \infty$ (the 
spherical limit). We consider the system confined in one direction (we will
call this the z - direction) within an extension L and the boundary 
conditions will be taken to be of Dirichlet type , i.e., $\phi_i =0$ at $z=0$
and $z=L$ . The expansion in Fourier modes for $\phi_i(\vec{r})$ is
\bea
\phi_i(\vec{r},z)= \frac{1}{\sqrt{L}} \int \frac{d^{D-1} k}{(2 \pi)^{D-1}}
\:\sum_{n} \phi_{in}(\vec{k})\exp{i \vec{k}.\vec{R}}\: sin\: \frac{n\pi z}{L}
\eea 
The specific heat is given by the correlation function
\ber
C_p &=& \frac{1}{AL} \int d^{D-1}r_1 d^{D-1}r_2\:dz_1\:dz_2 <\sum_{i}\phi_i^2
        (\vec{r_1},z_1) \sum_{j}\phi_j^2(\vec{r_2},z_2)>\nonumber \\
    &=& \frac{1}{4L}\sum_{n_1,n_2}\int \frac{d^{D-1}p_1}{(2\pi)^{D-1}}\:\:
	    \frac{d^{D-1}p_2}{(2\pi)^{D-1}}<\sum_{i}\phi_{in_1}(\vec{p_1})\:
		\phi_{in_1}(\vec{-p_1}) \nonumber \\ 
	& &	\sum_{j}\phi_{jn_2}(\vec{p_2})\:\phi_{jn_2}(\vec{-p_2})>
\eer
The free energy of Eq(2) written in terms of the fields $\phi_{in}$ because
\ber
F &=& \sum_{n}\int \frac{d^{D-1}k}{(2\pi)^{D-1}}\: [\:\frac{1}{2}(m^2+k^2+
      \frac{n^2\pi^2}{L^2})\:\phi_{in}(k)\:\phi_{in}(-k)\:] \nonumber \\ 
  &	& + \frac{\lambda}{4}\sum_{i}\sum_{j}
      \int \frac{d^{D-1}k_1}{(2\pi)^{D-1}}\:\:\frac{d^{D-1}k_2}
	  {(2\pi)^{D-1}}\:\:\frac{d^{D-1}k_3}{(2\pi)^{D-1}}\:\:dz\nonumber \\
  & & \phi_i(k_1,z)\:\phi_i(k_2,z)\:\phi_j(k_3,z)\:\phi_j(-k_1-k_2-k_3,z)
\eer
and the averaging shown in Eq(4) has to be done with this free energy 
functional.
\par
The Gaussian limit has to be disposed of first i.e. the limit where $\lambda
=0$ and the action is quadratic.                                
\ber
C_p^{Gaussian} &=& \frac{1}{4L}\:\sum_{n_1,n_2}\:\int\frac{d^{D-1}p_1}{(2\pi)
                   ^{D-1}}\:\frac{d^{D-1}p_2}{(2\pi)^{D-1}}\:<\phi_{i,n_1}
				   (\vec{p_1})\:\phi_{j,n_2}(\vec{p_2})>\nonumber \\
			& &<\phi_{i,n_1}(\vec{p_1})\:\phi_{j,n_2}(\vec{p_2})>\nonumber \\  
            &=&\frac{N}{2L}\:\sum_{n}\int \frac{d^{D-1}p}{(2\pi)^{D-1}}\:
               [\:<\phi_{i,n}(\vec{p})\:\phi_{i,n}(-\vec{p})\:]^2\nonumber \\
               &=&\frac{N}{2L}\:\sum_{n}\int \frac{d^{D-1}p}{(2\pi)^{D-1}}
                  \:\frac{1}{(p^2+m^2+\frac{n^2\pi^2}{L^2})}\nonumber \\
	           &=&\frac{NL^4}{2L}\int \frac{d^{D-1}p}{(2\pi)^{D-1}}\: 
			      [\:\:\frac{coth \sqrt{p^2L^2+m^2L^2}}{2(p^2L^2+m^2L^2)^{3/2}}
			   	 \nonumber \\
			   & &-\:\frac{1}{(p^2L^2+m^2L^2)^2}\:+\:\frac{cosech^2 \sqrt
			      {p^2L^2+m^2L^2}}{2(p^2L^2+m^2L^2)}\:]\nonumber \\
			   &=&\frac{N}{2} I_D(mL)
\eer			   	  
where
\ber
I_D(mL) &=& L^3\int \frac{d^{D-1}p}{(2\pi)^{D-1}}\:[\:\frac{coth 
            \sqrt{p^2L^2+m^2L^2}}{2(p^2L^2+m^2L^2)^{3/2}}\nonumber \\
		& &-\:\frac{1}{(p^2L^2+m^2L^2)^2}\:+\:\frac{cosech^2 \sqrt
			      {p^2L^2+m^2L^2}}{2(p^2L^2+m^2L^2)}\:]				   
\eer				 				 
For $L \gg m^{-1}$ this gives the usual gaussian limit answer $C_p \sim
m^{-(4-D)/2}$, while for $m\rightarrow 0$ at finite L, $C_p \sim L^{(4-D)/2}$.
The scaling function is provided by $I_D$.
\par
We now turn to the spherical limit. The propagator $<\phi_{in}(k)\:\phi_{in}
(-k)>$ has the structure $(M^2+k^2+\frac{n^2\pi^2}{L^2})^{-1}$ where M is the
renormalized mass dressed by the bubble [4] in the spherical limit. As usual
$M\sim (T-T_c)^{-\nu}$, where in the spherical limit $\nu=(D-2)^{-1}$. The 
specific heat graphs [4] for the spherical limit are shown in Fig.1. The first
graph (1a) corresponds to the gaussian limit. Each additional loop brings an
interaction of strength $N^{-1}$ and a combinatoric factor N (for large N).
Thus, each of the graphs (1b,1c,1d....etc) are the same order as the single 
loop and the sum defines the spherical limit. For $L\rightarrow\infty$,
the contributions factor and the successive contribution factors are $I^2$ (1b)
,$I^3$ (1c)....etc. The total specific heat is proportional to $\frac{I}{1+I}$
. In the case of finite geometry this simple geometric series does not obtain
as we show below.
\par 
To understand the complication, let us look at the two loop graph. The 
contribution from the graph is ,
\ber
  { }&=& -\frac{1}{L^3}\lambda\:\sum_{n_1,n_2}\int \frac {d^{D-1}p_1}
      {(2\pi)^{D-1}}\:\frac{d^{D-1}p_2}{(2\pi)^{D-1}}<\phi_{i,n_1}(p_1)\:\phi_
         {i,n_1}(-p_1)\int dz \:\sum_{s,t}\nonumber \\  
     & & \sum_{m_1,m_2,m_3,m_4}\phi_{s,m_1}(k_1)\:\phi_{s,m_2}(k_2)\:
         \phi_{t,m_3}(k_3)\:\phi_{t,m_4}(-k_1-k_2-k_3)\nonumber \\ 
     & & \: sin{\frac{m_1\pi z}{L}}\:sin{\frac{m_2\pi z}{L}}\: sin{\frac
         {m_3\pi z}{L}}\:sin{\frac{m_4\pi z}{L}}\phi_{j,n_2}(p_2)\:\phi_{j,n_2}
         (-p_2)>\nonumber \\ 	   
    &=&-\lambda \frac{N^2}{L^3}\int \frac {d^{D-1}p_1}{(2\pi)^{D-1}}\:
	\frac{d^{D-1}p_2}{(2\pi)^{D-1}}\:[\:<\phi_{i,n_1}(p_1)\:\phi_{i,m_1}(-p_1)>
	    \:]^{2}\nonumber \\
& &	   [\:<\phi_{j,n_2}(p_2)\phi_{j,m_2}(-p_2)>\:]^{2}\int dz \:sin^2 
       {\frac{m_1\pi z}{L}} \:sin^2 {\frac{m_2\pi z}{L}}\nonumber \\	   
   &=&-\frac{N}{4L^3} \sum_{n_1,n_2}\int \frac {d^{D-1}p_1}{(2\pi)^{D-1}}
 \:\frac{d^{D-1}p_2}{(2\pi)^{D-1}}\:[\:<\phi_{i,n_1}(p_1)\:\phi_{i,n_1}(-p_1)>
      \:]^{2}\nonumber \\
& &	   [\:<\phi_{j,n_2}(p_2)\:\phi_{j,n_2}(-p_2)>\:]^{2} \int dz\:(1+cos{\frac
       {2n_1\pi z}{L}})\:(1+cos{\frac{2n_2\pi z}{L}})\nonumber \\
   &=& -\frac{N}{4L^2}\sum_{n_1,n_2}\int \frac {d^{D-1}p_1}{(2\pi)^{D-1}}
 \frac{d^{D-1}p_2}{(2\pi)^{D-1}}\:[\:<\phi_{i,n_1}(p_1)\:\phi_{i,n_1}(-p_1)>\:]
       ^{2}\nonumber \\
& &	   \:[\:<\phi_{i,n_2}(p_2)\:\phi_{i,n_2}(-p_2)>\:]^{2} (\:1+\frac{1}{2}
	   \delta_{n_1,n_2}\:)\nonumber \\ 
&=& -\frac{N}{2}\:[\:I_D^2\:+\:\frac{1}{4L^2}\:\:\sum_{n}\:\int 
\frac{d^{D-1}p_1}{(2\pi)^{D-1}}\:\:\:\frac{d^{D-1}p_2}{(2\pi)^{D-1}}
\nonumber \\
& &\frac{1}{(p_1^2+M^2+\frac{n^2\pi^2}{L^2})^2}\:\:\:\:\frac{1}{(p_2^2+M^2+
\frac{n^2\pi^2}{L^2})^2}\:]\nonumber \\
   &=& -\frac{N}{2}\:[\:I_D^2+J_D^{(1)}\:] 
\eer
where
\bea
J_D^{(1)} = \frac{1}{4L^2}\:\:\sum_{n}\int\: \frac {d^{D-1}p_1}{(2\pi)^{D-1}}\:
\frac{d^{D-1}p_2}{(2\pi)^{D-1}}\:\:\frac{1}{(p_1^2+M^2+\frac{n^2\pi^2}{L^2})^2}
\nonumber \\
\frac{1}{(p_2^2+M^2+\frac{n^2\pi^2}{L^2})^2}
\eea
The difficulty is obvious from Eqn(8). The presence of the term $J_D^{(1)}$
means that the series for the specific heat is not a geometric series. The
three loop term analysed in a similar manner yields $\frac{N}{2}(I_D^3+2I_D
J_D^{(1)}+J_D^{(2)})$ , where 
\ber
J_D^{(m)}&=& \frac{1}{(2L)^{2m}}\:\sum_{n}\:\int\:\frac {d^{D-1}p_1}{(2\pi)^
{D-1}}\:\frac{d^{D-1}p_2}{(2\pi)^{D-1}}\:...\:\frac{d^{D-1}p_{m+1}}{(2\pi)^
            {D-1}}\nonumber \\
		& &\frac{1}{(p_1^2+M^2+\frac{n^2\pi^2}{L^2})^2}\frac{1}
            {(p_2^2+M^2+\frac{n^2\pi^2}{L^2})^2}\:...\frac{1}{(p_{m+1}^2+M^2
            +\frac{n^2\pi^2}{L^2})^2}\nonumber \\
		& &		
\eer
The four loop calculation is $-\frac{N}{2}\:[\:I_D^4\:+\:3I_D^2 J_D^{(1)}\:+\:
2I_D J_D^{(2)}\:+\:{J_D^{(1)}}^2\:+\:J_D^{(3)}\:]$. Similarly,the five loop 
term is $\frac{N}{2}\:[\:I_D^5\:+\:4I_D^3J_D^{(1)}\:+\:3I_D {J_D^{(1)}}^2\:+\:
2I_D J_D^{(3)}\:+\:2J_D^{(1)}J_D^{(2)}\:+\:J_D^{(4)}\:]$.The contribution from
the six loop term is $-\frac{N}{2}\:[\:I_D^6\:+\:5I_D^4 J_D^{(1)}\:+\:
2I_D^3 J_D^{(2)}\:+\:6I_D^2 {J_D^{(1)}}^2\:+\:6I_D J_D^{(1)}J_D^{(2)}\:+\:
{J_D^{(1)}}^3\:+\:3I_D^2 J_D^{(3)}\:+\:2I_D J_D^{(4)}\:+\:2J_D^{(1)} J_D^{(3)}
\:+\:{J_D^{(2)}}^2\:+\:J_D^{(5)}\:]$ and so on.
\par
The pattern is now clear. The specific heat is obtained as 
\ber 
C &=& \frac{N}{2}\:\{\:[I_D - I_D^2 + I_D^3 +.....] - [ J_D^{(1)} - J_D^{(2)} 
      + J_D^{(3)}-..][ 1 - 2I_D + 3I_D^2 - ..]\nonumber \\
& & - [ J_D^{(1)} - J_D^{(2)} + J_D^{(3)} - ..]^2 [ 1 - 3I_D + 6I_D^2-..] - 
      [ J_D^{(1)} - J_D^{(2)}\nonumber \\
& &	  + J_D^{(3)} -..]^3 [ 1 - 4I_D + 10I_D^2-..]....\}\nonumber \\ 
  &=& \frac{N}{2}\:\{\:\frac{I_D}{1+I_D} - \frac{\sum_{m}J_D^{(m)}}{(1+I_D)^2} 
      -\frac{(\sum_{m}J_D^{(m)})^2}{(1+I_D)^3}-\frac{(\sum_{m}J_D^{(m)})^3}
      {(1+I_D)^4} -..\}\nonumber \\
  &=& \frac{N}{2}\:\{\frac{I_D}{1+I_D} - \frac{\sum_{m}J_D^{(m)}}{(1+I_D)^2}
      [1 + \frac{\sum_{m}J_D^{(m)}}{(1+I_D)} + (\frac{\sum_{m}J_D^{(m)}}
      {(1+I_D)})^2 + ..\}\nonumber \\
  &=& \frac{N}{2}\:\{\frac{I_D}{1+I_D} - \frac{\sum_{m}J_D^{(m)}}{(1+I_D)^2}
      \frac{1}{1 - \frac{\sum_{m}J_D^{(m)}}{(1+I_D)}}\:\}\nonumber \\
  &=& \frac{N}{2}\frac{I_D -\sum_{m}J_D^{(m)}}{1 + I_D -\sum_{m}J_D^{(m)}}
\eer
This is the final answer for any dimension. The first part of Eqn(10), where
we generalize the pattern can be proven by induction [12] . We assume that 
this form is consistent with the form obtained after a m-loop calculation and 
prove in a long but straightforward fashion that it holds for the $(m+1)$ loop
as well. We now write down the explicit expression for $I_D$ and $J_D^{(m)}$
in $D=3$ .
\bea
I_3 = \frac{1}{2M}\:[\:Coth ML -\frac{1}{ML}\:]
\eea
For $L\rightarrow \infty$ , $I_3=M^{-1}$ as expected while for $M\rightarrow
0$ , $I_3=\frac{L}{3}$ in accordance with finite size scaling approximations.
Straightforward algebra yields
\bea 
J_3^{(1)} = -\frac{1}{4L^2}\:\frac{\partial I_3}{\partial M^2}
\eea
and for general m
\bea
J_3^{(m)} = \frac{(-)^m}{2^(2m) m!}\:\frac{1}{L^{2m}}\:\frac{\partial^m I_3}
            {\partial M^{2m}}
\eea
which allows us to write Eqn(10) as 
\bea
C = \frac{L F(\sqrt{M^2 L^2+\frac{1}{4}})}{1+L F(\sqrt{M^2 L^2+\frac{1}{4}})}
\eea
where
\bea
F(X) = \frac{1}{X}\:[\:Coth X -\frac{1}{X}\:]
\eea
The spherical limit scaling function is contained in Eqns(14) and (15) - 
the central result of our paper. 
\par
We now show how the above equations can be used in the practical situation of
liquid $He^{4}$ near the superfluid transition in $D=3$ . The experimental 
specific heat is known to be almost logarithmic, while in the spherical limit 
in $D=3$ , $\frac{\alpha}{\nu}=-1$ . We can split off a background part from 
Eqn(14) by writing
\ber 
C &=& 1 - \frac{1}{1+L F(\sqrt{M^2 L^2+\frac{1}{4}})}\nonumber \\
  &\simeq&  1 - \frac{1}{L F(\sqrt{M^2 L^2+\frac{1}{4}})}
\eer
The second step follows from the fact that in the critical region the 
contribution $L F(\sqrt{M^2 L^2+\frac{1}{4}})$ is expected to be large. It is 
the factor $\frac{1}{L F(\sqrt{M^2 L^2+\frac{1}{4}})}$ which describes the 
scaling function with $\frac{\alpha}{\nu}=-1$ . If we want to write the answer
in terms of $\frac{\alpha}{\nu}$ , then the specific heat is $[\:L F(\sqrt{X^2+
\frac{1}{4}})\:]^{\frac{\alpha}{\nu}}$ and for $\frac{\alpha}{\nu}\rightarrow 0$
as is the experimental situation, we can write 
\ber
C(M,L) &=& C_0\: ln\:[\:L F(\sqrt{M^2 L^2+\frac{1}{4}})\:] + constant \nonumber \\ 
       &=& C_0\: ln\:[\:\frac{\Lambda L}{\sqrt{M^2 L^2+\frac{1}{4}}}(Coth
           \sqrt{M^2 L^2+\frac{1}{4}}-\frac{1}{\sqrt{M^2 L^2+\frac{1}{4}}})\:]
	   	   \nonumber \\
	   & &{ } 	   
\eer
where $\Lambda = \kappa_0 t_0^{\nu}$. For $L\rightarrow \infty$ , the bulk 
specific heat
\ber
C &=& C_0\: ln\: [\:\frac{\Lambda}{M}\:]\nonumber \\
  &=& C_0\: ln\: [\:\frac{\Lambda}{\kappa_0 t^{\nu}}\:]\nonumber \\
  &=& \nu\: C_0\: ln[\:\frac{t_0}{t}\:]
\eer
The experimentally measured bulk value has this form [13] and we identify 
$\nu C_0 = 5.3$ and $t_0=\frac{1}{4}$ . The value at $M=0$ is 
$C_0\:ln\:[2\Lambda L(Coth\frac{1}{2}-2)]$. Subtracting the $C_0\:ln\:
\Lambda L$ from Eqn(18) to obtain a pure scaling part
\ber
\Delta C(M,L) &=& C(M,L)-C_0\:ln\:\Lambda L\nonumber \\
&=& C_0\:ln\:[\:\frac{(Coth\sqrt{X^2 + \frac{1}{4}}-\frac{1}{\sqrt{X^2 + 
    \frac{1}{4}}})}{\sqrt{X^2 + \frac{1}{4}}}\:]
\eer
where $X=ML$ . We show this plot in Fig(2) and the data of Mehta and Gasparini
[14]. The agreement is the proof of the practical utility.  \\\\

$ {\bf {Acknowledgments}} $  \\\\\
One of the authors (SB) would like to thank the C.S.I.R, India, for providing 
partial financial support and Dr.Manabesh Bhattacharya for his help and 
encouragement.

\newpage
$ \bf {Figure\:\:caption} $ \\\\

Fig.1. Specific heat graphs in the spherical limit are shown.\\
Fig.2. $\Delta C= C(M,L)- C_0\:ln\:\Lambda L$ ,Eqn(20), plotted against 
$(ML)^{1/\nu}$ . The solid curve refers to our theory and the data is taken 
from the experiment of Mehta and Gasparini [14]
\\


\newpage
\begin{thebibliography}{99}
\bibitem{1} T. H. Berlin and M. Kac, Phys. Rev {\bf{86}} 821 (1952)
\bibitem{2} R. Abe and S. Hikami, Prog. Theor. Phys {\bf{49}} 442 (1973)
\bibitem{3} R. A. Ferrell and D. J. Scalapino, Phys. Rev. Lett {\bf{29}}
 413 (1973)
\bibitem{4} S. K. Ma, Phys. Rev {\bf{A7}}  2172  (1973)
\bibitem{5} A. J. Bray, Phys. Rev. Lett {\bf{32}}  1413  (1974) 
\bibitem{6} P. W. Anderson in "Valence Fluctuations in Solids", ed.
L. M. Falikov et.al. (North Holland- Amsterdam)
\bibitem{7} E. Brezin, C. Itzykson, G. Parisi and J. B. Zuber, Comm. Math.
Phys. {\bf{59}}  35  (1978)
\bibitem{8} E. Witten, Physics Today, pp 38 , July (1980)
\bibitem{9} A. J. Bray and M. A. Moore, Phys. Rev. Lett.{\bf{38}}  735 (1977)
\bibitem{10}T. Vojta, Phys. Rev. {\bf{B53}}  710  (1996)
\bibitem{11}J. Ye, S. Sachdev and N. Read , Phys. Rev. Lett {\bf{70}} 4011
(1993)
\bibitem{12}S. Bhattacharyya, Ph.D Thesis Jadavpur University. (1998)
\bibitem{13}L. S. Goldner and G. Ahlers, Phys. Rev. {\bf{B45}} 13129  (1992) 
\bibitem{14}S. Mehta and F. M. Gasparini, Phys. Rev. Lett {\bf{78}} 2596
(1997)
\end{thebibliography}
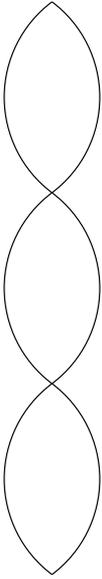
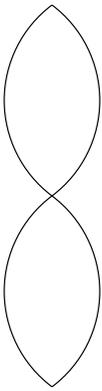
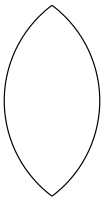
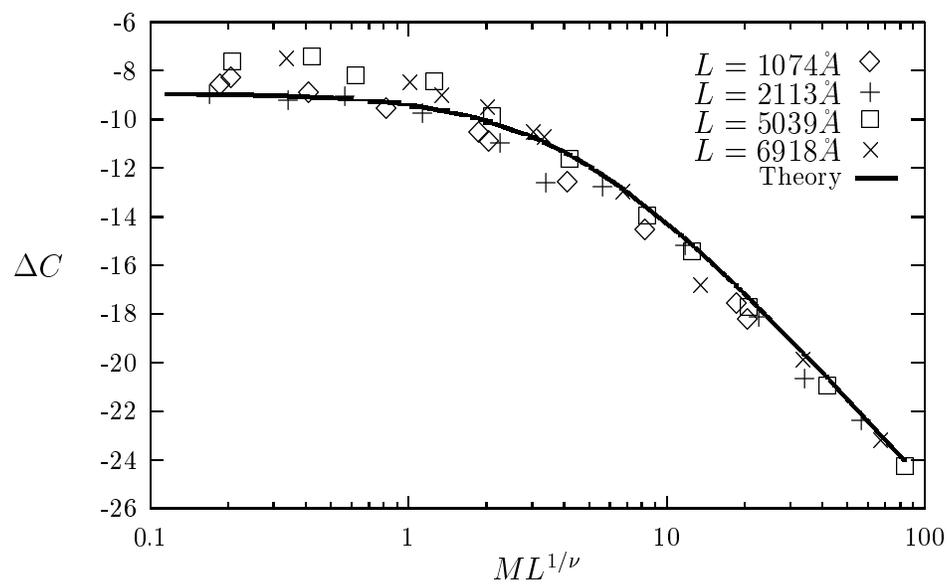
\end{document}